\documentclass[prl,preprint,showpcs,amsmath,amssymb]{revtex4}
\usepackage{graphicx}
\usepackage{dcolumn}
\usepackage{bm}

\begin{document}


\title{Ordering of small particles in one-dimensional coherent structures \\ by time-periodic flows \\ }

\author{D.O. Pushkin, D.E. Melnikov, and V.M. Shevtsova }
\affiliation{Microgravity Reaserch Center, Univetsit{\'e} Libre de Bruxelles, CP-165/162, B-1050, Bruxelles, Belgium}

\date{\today}
           
\begin{abstract}
Small particles transported by a fluid medium do not necessarily have to follow the flow.  We show that for a wide class of time-periodic incompressible flows inertial particles have a tendency to spontaneously align in one-dimensional dynamic coherent structures. This effect may take place for particles so small that often they would be expected to behave as passive tracers and be used in PIV measurement technique. We link the particle tendency to form one-dimensional structures to the nonlinear phenomenon of phase locking. We propose that this general mechanism is, in particular, responsible for the enigmatic formation of the `particle accumulation structures' discovered experimentally in thermocapillary flows more than a decade ago and unexplained until now.

\end{abstract}


\maketitle

\newpage

Ordering and transport of small particles suspended in an incompressible fluid medium is both fundamentally interesting and important in an array of natural and technological processes. Their examples include manipulation and segregation of biological and synthetic microparticles~\cite{CarloToner}, concentration and transport of pollutants, nutrients, and plankton by oceanic currents and of aerosols and droplets by atmospheric currents~\cite{Seinfeld98}. 

When interparticle interactions are negligible, small particles are often expected to behave as passive tracers when their Stokes number, defined as the ratio of the particle response time to the characteristic flow time scale, is much less than unity. On the other hand, it has been well-known in the fluid mechanical literature that small but finite sized particles with densities different from that of the surrounding fluid may deviate from the fluid trajectories. These deviations, accrued over time, lead to non-conservative trajectories in otherwise volume-preserving flows. This fundamental effect may have profound consequences. In the past it has been numerously invoked to explain segregation and unmixing of heavy particles~\cite{Max87}. Recently a strong theoretical interest to this effect resulted in predicting the possibility of permanent suspension of particles in flows under gravity~\cite{Gravity}, formation of three-dimensional particulate coherent structures in hurricanes~\cite{SaH}, trapping of aerosol particles in open chaotic flows~\cite{VilelaMotter07} and in a chaotic wake~\cite{BTT02}. A systematic treatment of the coherent structures arising due to inertia effects along the lines of dynamical systems was undertaken recently~\cite{SaH}.    

\begin{figure}
\includegraphics[scale=0.2]{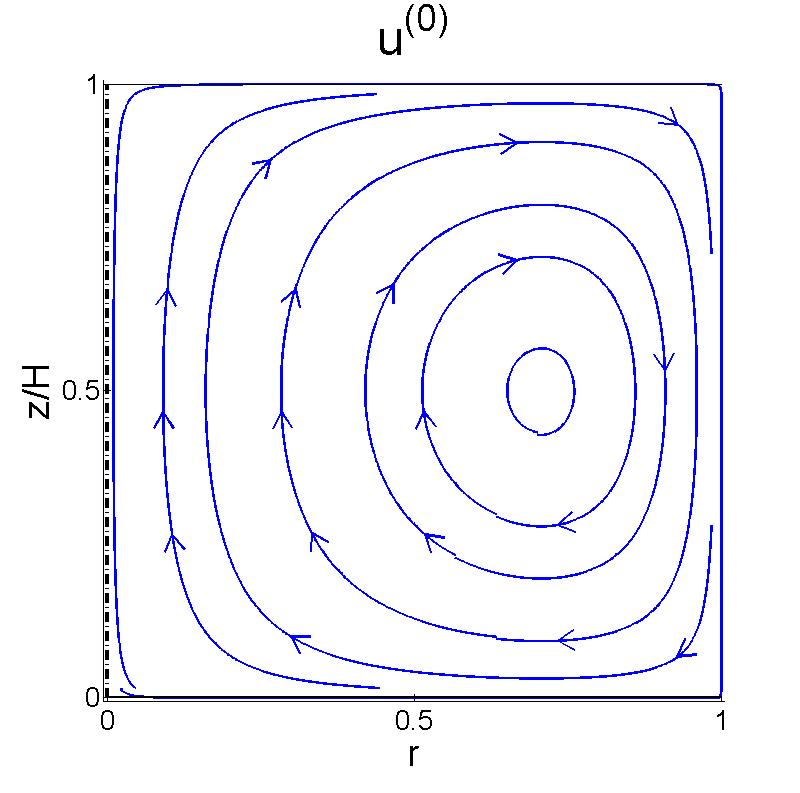}
\includegraphics[scale=0.2]{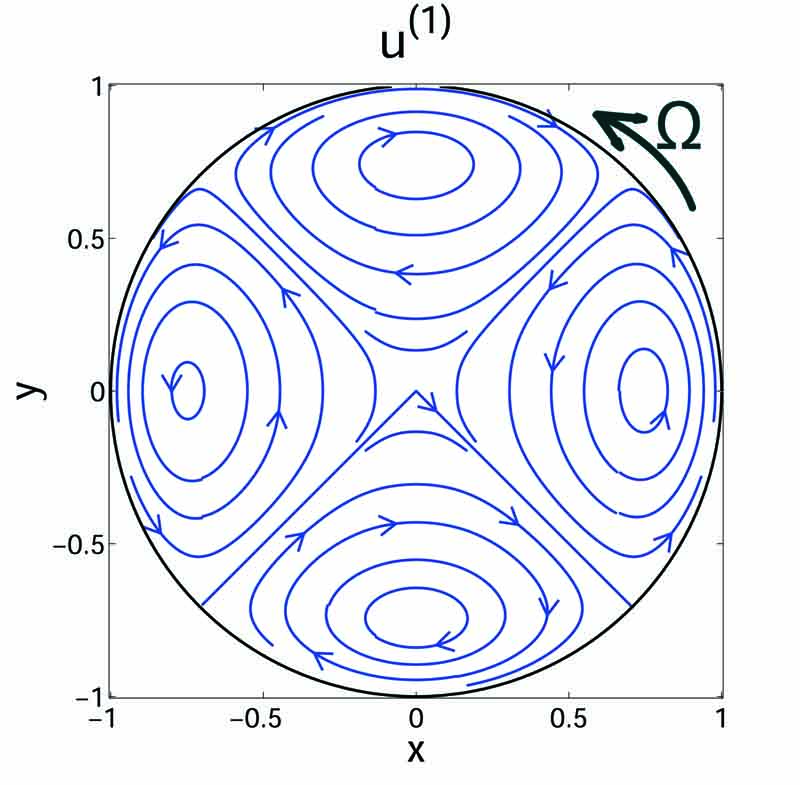}
\caption{\label{fig:Flow}
Sketch of a time-periodic flow with an azimuthal traveling wave. Here the wave mode $m=2$. Such flows often occur  in the cylindrical geometry as a result of an instability~\cite{HM75,Lopez04,PSS83}. In the experiments of Schwabe et al. the flow was induced by the Marangoni effect\cite{ScrivenSternling60} in a drop placed between two rods, with the top rod heated.
}
\end{figure}

In this Letter we report a new type of ordering of inertial particles that results in formation of {\it one-dimensional} dynamical particulate coherent structures. It is rather surprising that while this effect is generic for a class of widely encountered time-periodic flows, it has apparently not been analyzed previously. It is even more surprising that the inertia-driven self-assembly of particles into one-dimensional continuous lines was actually  observed in experiments on thermocapillary flows more than a decade ago by Schwabe et al.~\cite{SHF96} but despite extensive experimental studies~\cite{TanakaSchwabe06,SMT07} has remained unexplained. Our results were first reported at~\cite{PMS10}.

The fluid flows we consider occur in the cylindrical geometry. They can be represented as superpositions of a steady toroidal vortex $\bm{u^{(0)}}(r,z)$ and an oscillatory wave traveling azimuthally $\bm{u^{(1)}}(\phi - \Omega t, r, z)$, see Fig.~\ref{fig:Flow}. Here $(\phi,r,z)$ are the cylindrical coordinates, $t$ is time, and $\Omega$ is the wave angular velocity. Such flows often emerge as a result of an instability, e.g. the convective instability in a rotating annulus that had been suggested as a model for atmospheric circulation~\cite{HM75} or the instability in a stationary open cylinder driven by rotating endwall~\cite{Lopez04}. The ubiquity and practical importance of such flows is the reason why a number of studies have been devoted to them, particularly with the focus on chaotic advection and mixing~\cite{AlvarezShinbrotMuzzio02}. Here, however, our focus is different. 

\begin{figure}
\includegraphics[scale=0.35]{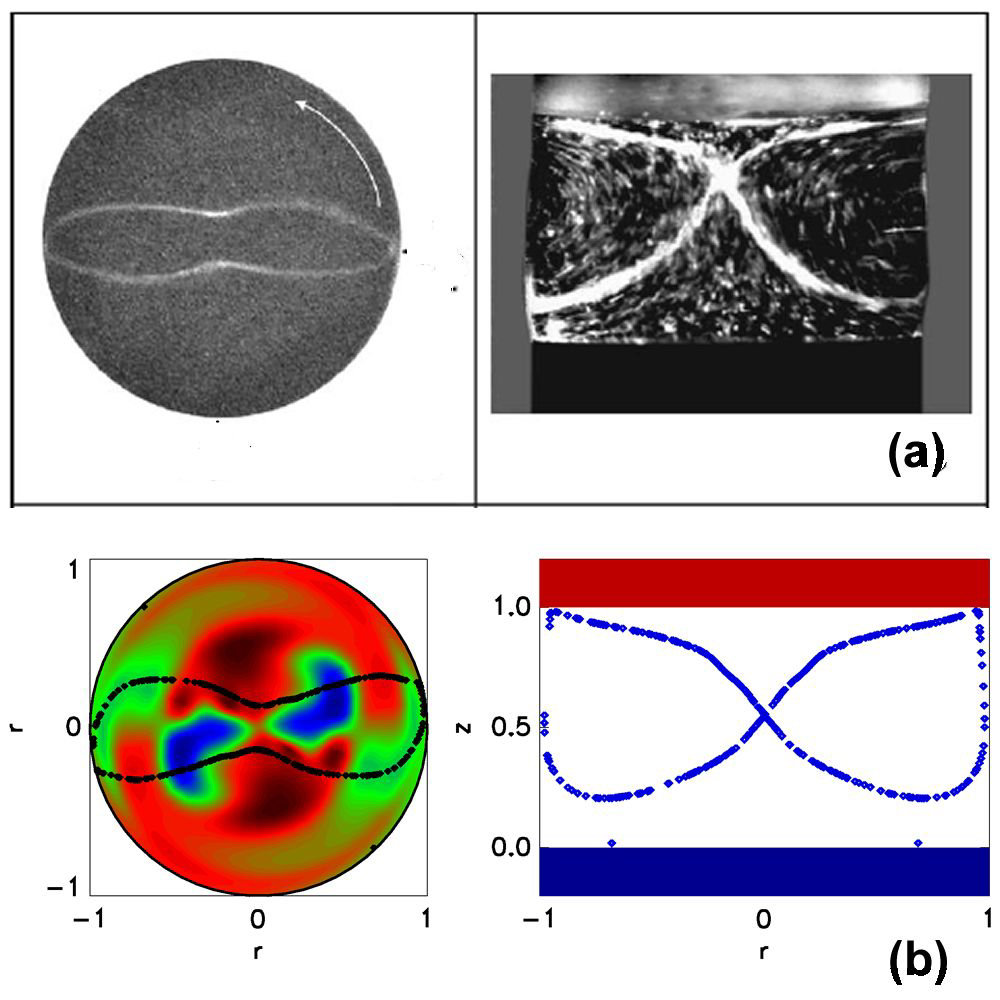}
\caption{\label{fig:PAS} Self-assembling particulate structures. (a), Azimuthal (left) and axial (right) snapshots of the spiral structures observed in the experiments of D. Schwabe et al.~\cite{SMT07}. The illuminated particles are seen as white specks. The closed spiral rotates in the direction denoted by the arrow. (b), Corresponding views of the structures obtained in our direct numerical simulations~\cite{PMS_LongPaper10}. The particle instantaneous positions are denoted by dots. On the left plot the particles positions are superimposed on the plot of the instantaneous temperature field taken at the cylinder mid-plane. Here $m=2$. }
\end{figure}

In the experiments of Schwabe the instability was of thermocapillary origin~\cite{PSS83}: it develops when a drop of liquid is placed between two cylindrical rods, with the top rod heated. The flow, driven at the cylindrical liquid--gas interface by the Marangoni force~\cite{ScrivenSternling60}, undergoes a bifurcation from a steady axially-symmetric vortex to an oscillatory regime with a traveling wave when the temperature difference between the rods is large enough. That particles will self-assemble in this flow was discovered when the experimentalists admixed particles, having sizes of tens of microns, in order to study the flow. To much surprise, they discovered that under certain conditions such small particles defied the fate of passive tracers and aligned themselves in an ordered spiral structure (see Fig.~\ref{fig:PAS}(a) and Supplementary Movie 1~\cite{SupplMat}). The spiral was closed, rather symmetric and rotated around the axis with no change of shape. It was dubbed `PAS' for `particle accumulation structure'~\cite{SHF96}. Notably, the angular frequency of rotation of the spiral was found equal to that of the wave. In other experiments particulate spirals of various shapes were observed both on the ground and in microgravity for small particles of different sizes, shapes, and densities~\cite{TanakaSchwabe06,SMT07}. However, the coherent structures were observed only for certain parameter ranges.

As the convective time-based Stokes number $St = 2/9 \left( a/L \right)^2 Re \sim 10^{-4} \div 10^{-3} \ll 1$, the particle dynamics is dominated by the viscous drag force. Here $a$ is the particle size, $L$ is the characteristic flow length scale, $Re=UL/\nu$ is the Reynolds number, $U$ is the characteristic flow speed, and $\nu$ is the fluid kinematic viscosity. In previous experimental studies of the particle self-ordering researchers looked for additional forces and accounted for flow features specific to thermocapillary and free-surface flows~\cite{SMT07}. However, we claim that this phenomenon is generic for the class of volume-preserving flows defined above, and that an interplay of the particle inertia and the viscous drag force alone may cause the ordering. Therefore we anticipate this effect in such periodic flows, which abound in nature. Their further examples include flows in laminar stirred tanks~\cite{AlvarezShinbrotMuzzio02}, pyroclastic surges, and microfluidic flows ~\cite{BB02}.
The arising singular spatial distributions of particles may have profound consequences for particle aggregation and transport; understanding and control of the arising structures may be important as a technological tool.

The starting point of our analysis is the reduced version of the Maxey-Riley equation of particle motion~\cite{MREq}, which can be written as
\begin{eqnarray}
\hskip 4mm \left( \rho+ \frac12 \right) {d \bm{v} \over d t} &=&  \frac32 {D \bm{u} \over D t} + \frac{1}{\tau} (\bm{u} -\bm{v}) + (\rho-1) \bm{g}.
\label{eq:MR1}
\end{eqnarray}
Here $t$ is time, $\bm v$ is the particle velocity, $\bm u$ is the fluid velocity, $\rho$ is the ratio of the particle to the fluid density, $\tau$ is the particle relaxation time, and $\bf g$ is gravity. The material derivative ${D \over D t}$ is taken moving with the fluid velocity at the current location of the particle. By the Stokes law, $\tau= (2/9) a^2 / \nu$. This equation describes the Newtonian dynamics of small particles dominated by the inertia (including the added mass effect) and viscous drag forces. It neglects a reverse influence of particles on the flow and particle--particle interactions and is valid when $a/L \ll 1$ and $U a/ \nu \ll 1$. (While in an actual physical system particles may experience other types of hydrodynamic forces, our goal is to demonstrate that even the current `minimal' model can produce the effect.)
 
We performed numerical simulations, in which the particle dynamics governed by (\ref{eq:MR1}) was coupled to the full system of the Navier-Stokes equations describing the thermocapillary fluid flow~\cite{ShevMelNep09,MelPushShev10}. We find, in physically realistic regimes, that particles assemble in dynamic spirals that closely resemble the experimental results (Fig.~\ref{fig:PAS}(b) and Supplementary Movie 2~\cite{SupplMat}).
Besides, we observed formation of particulate spirals that have the number of turns $l$ different from the wave mode $m$ (also integer due to the cylindrical geometry) \cite{PMS_LongPaper10}. Similar to experimental findings, each coherent structure is robust in a limited range of governing parameters. As soon as a parameter leaves the range, the coherent structure will disperse. We will come back to these features below. The major lesson learned from the simulations is the basic fact that a mere interplay of the inertia and viscous forces acting on individual particles can lead to the particle ordering.

\begin{figure}[t]
\flushleft
\centering
\begin{center}
\includegraphics[scale=0.18]{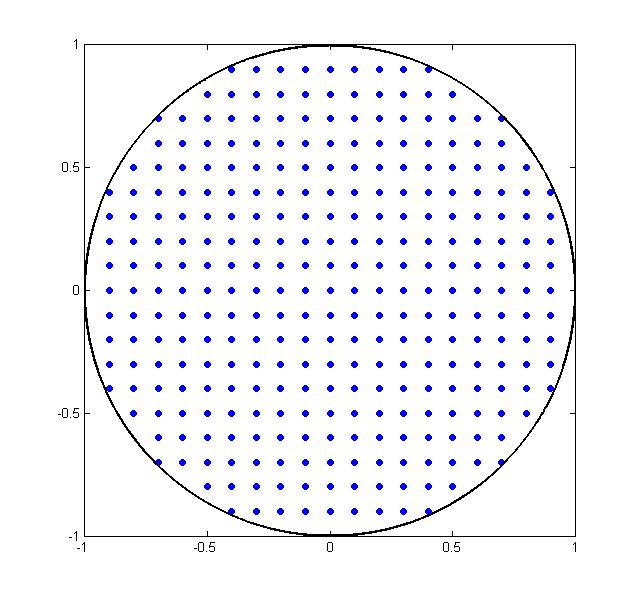}
\includegraphics[scale=0.18]{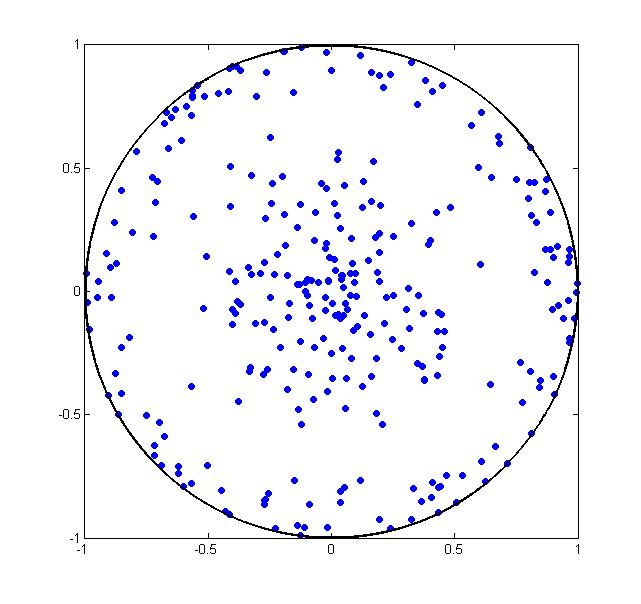}
\vskip 1mm
\includegraphics[scale=0.18]{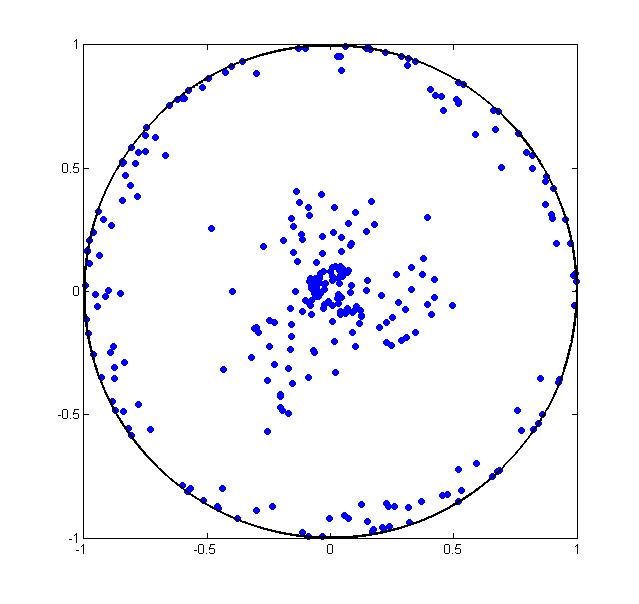}
\includegraphics[scale=0.18]{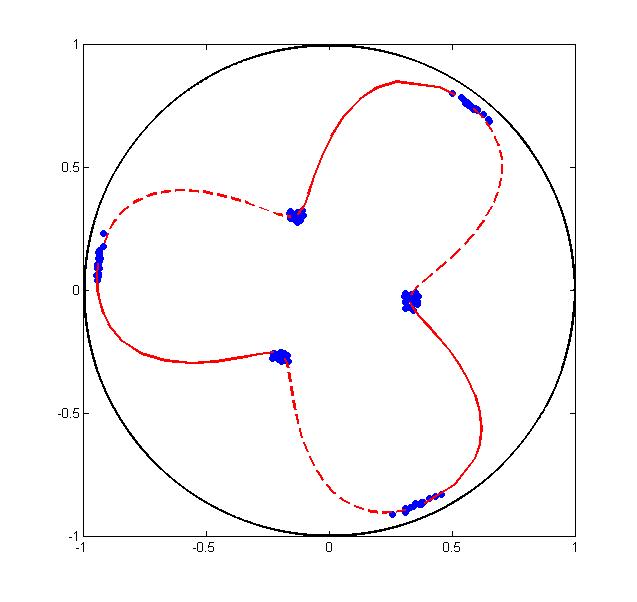}
\end{center}
\caption[]{Distribution of particles in the equatorial plane rotating with the wave after $n$ iterations of the Poincare map. (a) n=0, (b) n=150, (c) n=300, (d) n=3000. Here $m=3$, $H=1$, $b=0.3$, $\Omega=0.5$, and $\tau (\rho-1)=10^{-3}$ (see the second term in Eq.~\ref{eq:MR2}). In the physical space this picture corresponds to formation of a two-dimensional toroidal coherent structure that later transforms into a one-dimensional closed spiral. The projection of the spiral is showed in panel (d) by the red line. The physical formation time is about $10^3$.}
\label{fig:Poincare}
\end{figure}

At this point one could suppose that the effect is specific to the thermocapillary flow. In order to demonstrate that that this is not the case, we study particle ordering in a generic analytical model of a flow with a rotating wave. For a cylinder of unit radius we assume:	
\begin{subequations}
\label{eq:Flow}
\begin{eqnarray}
\bm{u}(t,\phi,r,z)&=&\bm{u^{(0)}}(r,z) + b \bm{u^{(1)}}(\phi - \Omega t, r, z) ,
\end{eqnarray}
\begin{eqnarray}
\label{eq:Flow0}
u^{(0)}_r(r,z)&=&-{\pi \over 2H}\left(r-r^3\right)\cos{ {\pi z \over H} },\nonumber \\ 
u^{(0)}_z(r,z)&=&-2\left( r^2- 1/2 \right)\sin {\pi z \over H},
\end{eqnarray}
\begin{eqnarray}
\label{eq:Flow1}
u^{(1)}_\phi(\phi,r,z)&=& - \left( 3r^2-4r^3 \right) \sin \left(m \phi \right) \sin{ {\pi z \over H} }, \nonumber \\ 
u^{(1)}_r(\phi,r,z)&=&   \left( r^2-r^3 \right) m \cos \left(m \phi \right) \sin{ {\pi z \over H} }.
\end{eqnarray}
\end{subequations}
\noindent 
Here $H$ is the cylinder height, $m$ is the (integer) wavenumber, and $b$ is the wave amplitude.  The equations (\ref{eq:Flow0}) mimic the Poiseuille flow in a finite cylinder, and the equations (\ref{eq:Flow1}) -- a rotating wave. The flow is scaled by $U$, so that the maximum value of $u_z$ is
$1$ at $z = H/2$ and $r = 0,$ or $1$. This model is not intended to be a rigorous description of the flow observed in experiments or numerical simulations; rather, it is a phenomenological model that captures the essential features of flows with rotating waves. It satisfies impermeability boundary conditions. A model satisfying no-slip conditions on the solid walls and zero axial vorticity component on the free surface would be significantly more complicated. Nevertherless, this simplified model successfully reproduces the principle features of the particle ordering, as is showed below. 

Instead of solving (\ref{eq:MR1}), we deal with the `inertial equation'~\cite{SaH}, obtained as the first order approximation of (\ref{eq:MR1}) in $St$:
\begin{eqnarray}
\label{eq:MR2}
\bm{v}(\bm{x})=\bm{u} - \tau (\rho-1) \left( {D \bm{u} \over D t} - \bm{g} \right) + O(St^2).
\end{eqnarray}
\noindent This equation describes particle motion as advection plus a perturbation due to the inertial effects. Since $\nabla \cdot \bm{u} = 0$, it is the nonzero divergence of the second term that must be responsible for the phase volume changes accompanying formation of the accumulation structures. 
The chief advantage of using (\ref{eq:MR2}) instead of (\ref{eq:MR1}) lies in the reduction of the particle phase space from six to three dimensions (in the frame rotating with the wave). Then from the viewpoint of dynamical systems the coherent structures are attractors in the three-dimensional space and can be readily studied by means of the Poincare section. We define the latter as the instances when particles cross the equatorial plane $z=1/2$ (in the reference frame rotating with the wave).

\begin{figure}
\flushleft
\centering
\includegraphics[scale=0.07]{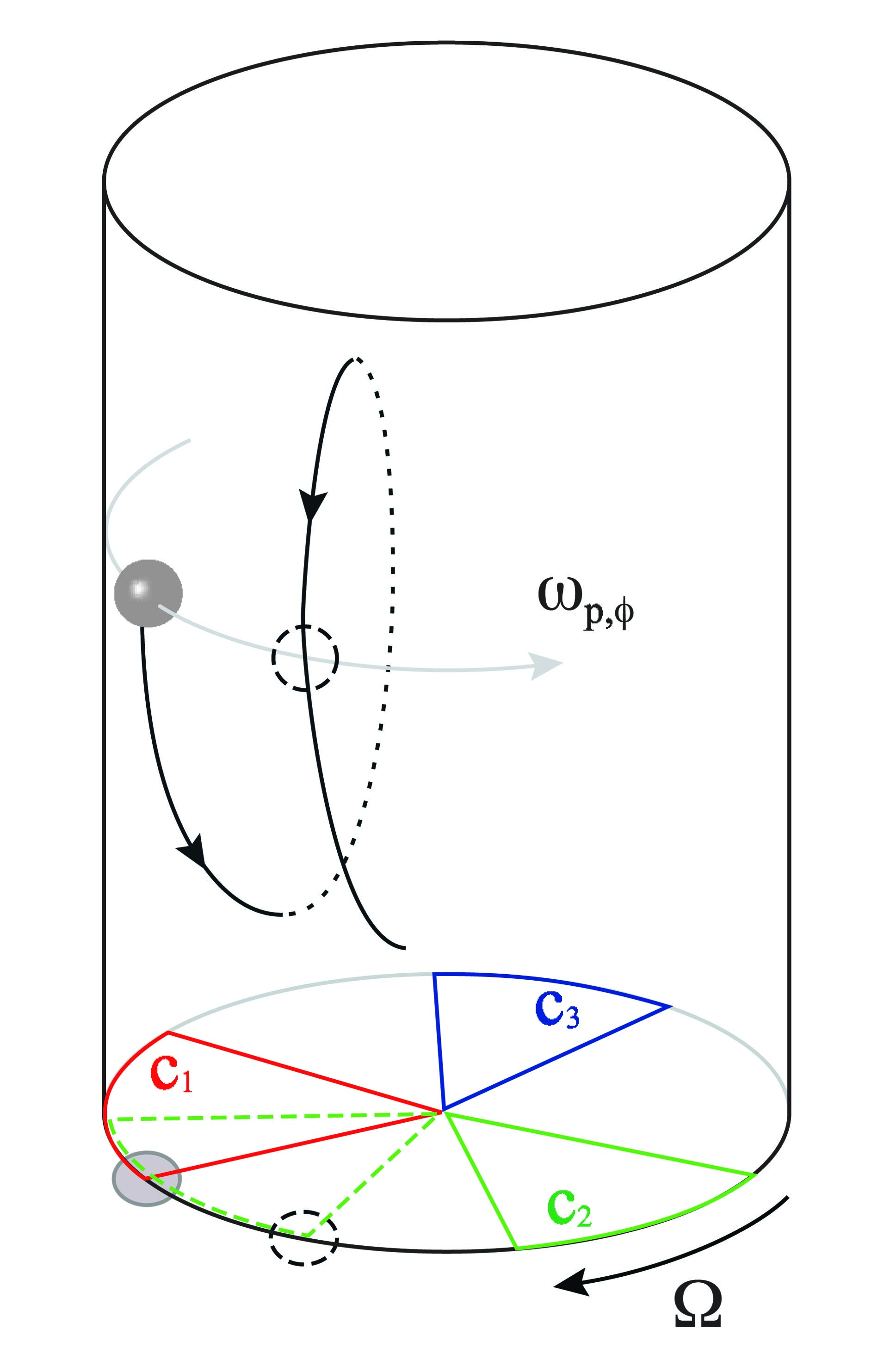}
\includegraphics[scale=0.05]{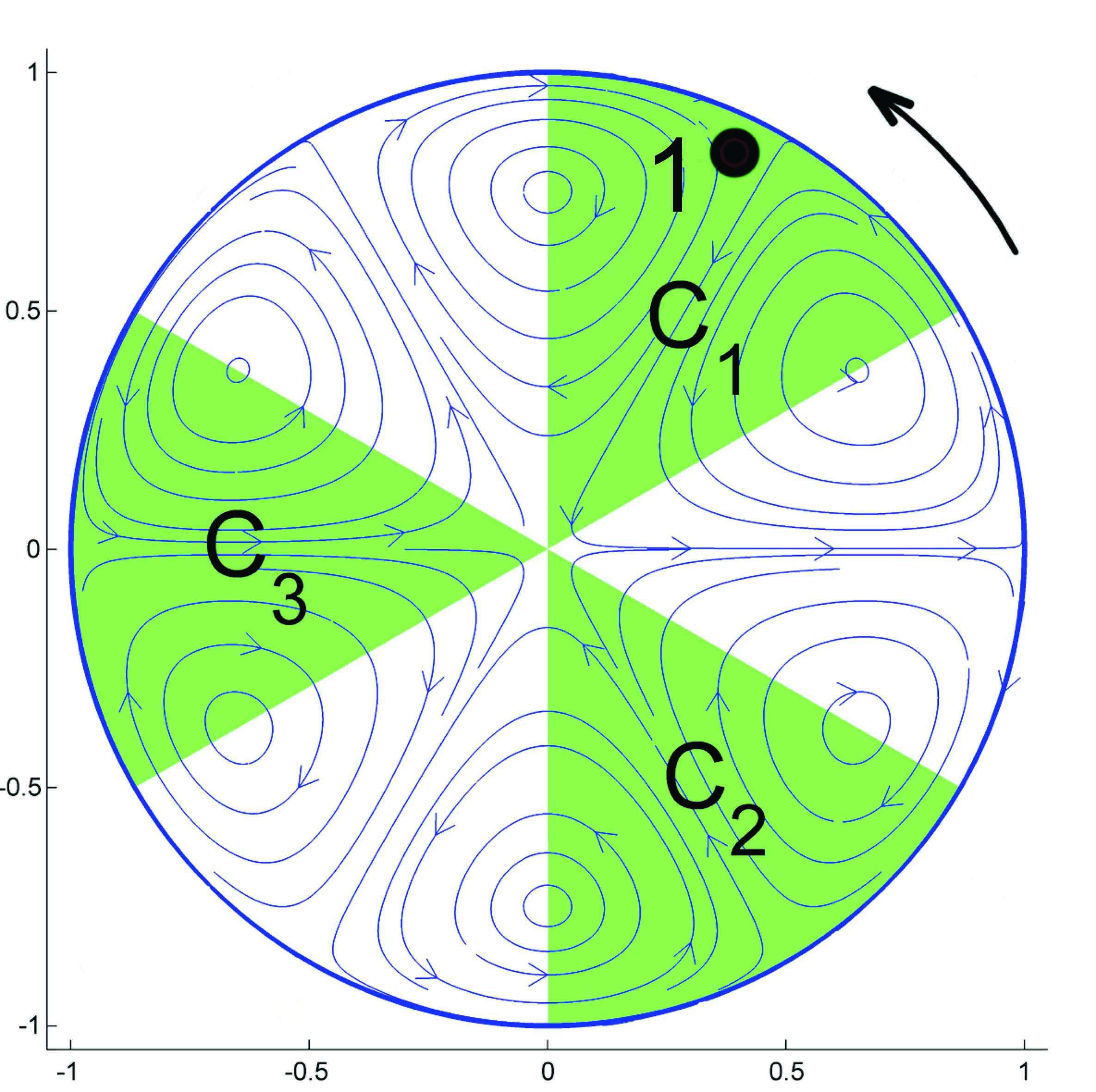}
\includegraphics[scale=0.05]{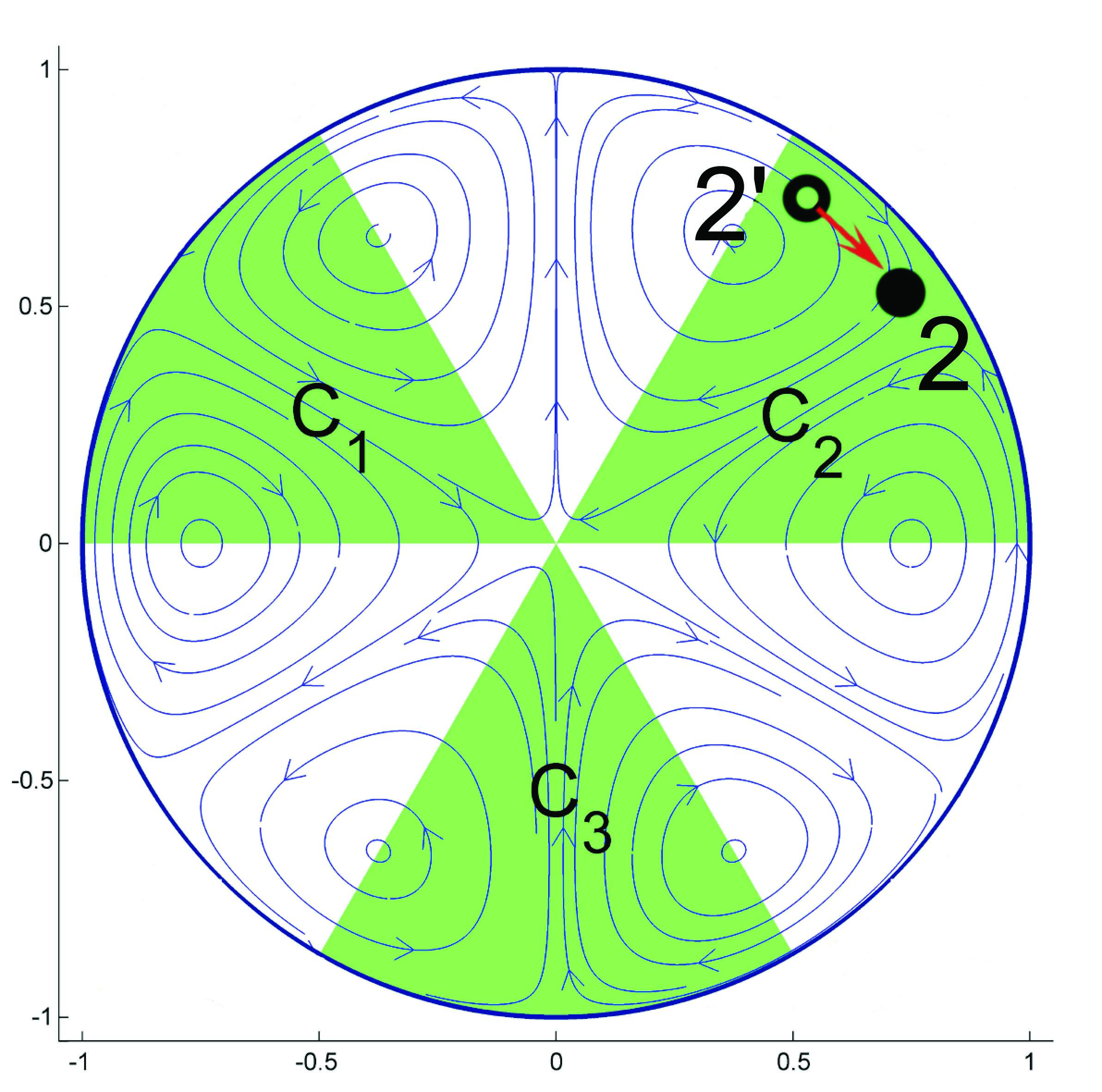}
\caption[]{Synchronization of a particle and the wave. Here $m=3$. (a) Schematic trajectory of an individual particle in the laboratory frame. (b),(c) The  equatorial plane in two instances of time separated by the particle turn-over time $T_p$. The phase locking means the particle phase relative to the wave is the same for the consecutive locations `$1$' and `$2$'. If the particle gets out of phase, `$2'$', interaction with the flow structure will adjust the particle azimuthal drift (red arrow) to resume the phase.}
\label{fig:PAS_synch}
\end{figure}

The model (\ref{eq:Flow}), (\ref{eq:MR2}) contains five parameters $H, m, a, \Omega,$ and $(\rho-1)\tau$, and can be studied numerically. At this point our goal is two-fold: first, to show that the model may account for the spontaneous formation of one-dimensional structures from the initially three-dimensional particle distributions for a realistic choice of parameters. Second, to demonstrate that it requires {\it no fine-tuning} of the parameters, i.e. the structures are stable for certain parameter ranges. 

We choose $H=1, m=3$, and $b=0.3$ to account for the flow and the rotating wave characteristics observed in experiments and our numerical simulations, and $(\rho-1)\tau=10^{-3}$ to account for the particle properties. The main tunable parameter is $\Omega$ and for now we set it to $0.5$. Numerical solution of (\ref{eq:Flow}), (\ref{eq:MR2}) bears out formation of coherent structures similar to the ones observed in experiments~\cite{TanakaSchwabe06,SMT07} and direct numerical simulations~\cite{MelPushShev10}. We clearly observe that particles do not need to touch the fluid--gas interface in order to align.  Fig.~\ref{fig:Poincare} demonstrates that the coherent structures are stable fixed points of the Poincare map. It also shows that the particle ordering proceeds via two distinct steps: first, particles concentrate in the center and near the circumference of the Poincare section. Second, the cylindrical symmetry is broken and the particles are attracted to the stable fixed points. In the physical space these steps correspond to clustering of particles in two-dimensional toroidal coherent structures, and to transformation of the latter into one-dimensional closed spirals. The two processes have different characteristic times but take place simultaneously~\cite{PMS_LongPaper10}. (These steps of the particle self-ordering had been noted in experiments~\cite{SM10}.) As a result, the motion of an individual particle in the spiral gets synchronized with the wave, and the resulting structure appears to rotate along with the wave in the laboratory frame. 

\begin{figure}
\includegraphics[scale=0.20]{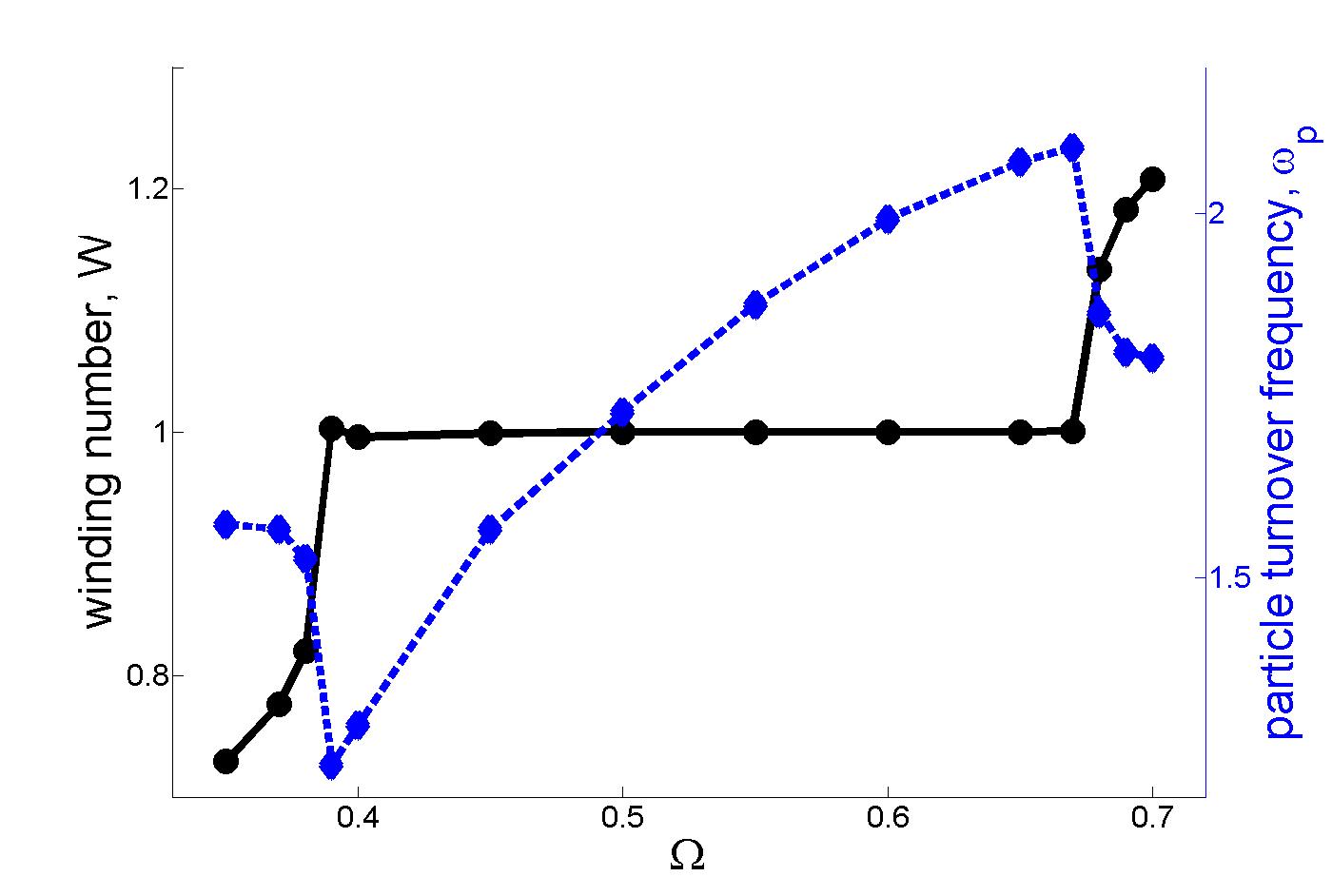}
\caption[]{The winding number plateau. The winding number $W$ (solid, black) and the particle turnover frequency $\omega_p$ (dashed, blue) vs. the wave angular velocity $\Omega$. While $\omega_p$ changes, $W$ remains constant for a range of values of $\Omega$. Constancy of winding number is a signature of phase locking~\cite{Ott93}.}
\label{fig:Winding}
\end{figure}

We find an explanation for this behavior in the nonlinear phenomenon of phase locking. Phase locking, alternatively called frequency locking or entrainment, typically occurs in dissipative systems with a weak interaction between modes with close frequencies~\cite{Ott93}. In our case the turn-over particle motion is synchronized with the rotating wave oscillations. The physical essence of the mechanism lies in the adjustment of the azimuthal particle displacement after every particle turn-over $\phi_{n+1}-\phi_n$ due to the inertial interaction with the wave, see Fig.~\ref{fig:PAS_synch}. Consequently, the particle azimuthal drift $\omega_{p,\phi}$ is also modified.

It is a signature of phase locking that the ratio of frequencies, called the winding number, remains a constant rational over a range of control parameters~\cite{Ott93}. In our case the winding number $W= \bar \Omega / \omega_p$, where $ \bar \Omega$ is the frequency of the wave oscillations experienced by the moving particle, and $\omega_p$ is the frequency of the particle turn-over motion. Fig.~\ref{fig:Winding} shows that while $\omega_p$ changes as $\Omega$ is varied over the range $[0.39,0.67]$, $W$ remains identically equal to $1$. The values of $W$ change abruptly outside the phase locked regime and the coherent structures loose stability. We find other rational values for the winding number $W=p:q$ for different parameter values and in certain regimes in direct numerical simulations~\cite{PMS_LongPaper10}. The latter equality can be re-written as the `resonant condition' $p \, \omega_p = q \, \bar \Omega $. (A similar `resonant condition' was noted in~\cite{SMT07}. Observe, however, that unlike resonance, phase locking does not require fine-tuning but provides it.) We note that $\bar \Omega = m ( \Omega - \omega_{p,\phi})$ (`angular Doppler shift'). Thus, the phase locking adjusts the particles azimuthal drift to maintain the `resonant condition.'  The three integers $p, q$, and $m$ fully determine the geometrical shape of the particulate spirals~\cite{PMS_LongPaper10}.

Synchronization due to phase locking is ubiquitous in nature. The present modeling suggests that PAS formation in thermocapillary flows is another instance of this general phenomenon and that formation of one-dimensional coherent particulate structures should be encountered in other oscillatory vortical flows when 1) the particle turnover motion is transversal to the direction of wave propagation and 2) the frequencies of the particle turnover motion are commensurate with the oscillation frequencies.



\begin{thebibliography}{34}
\expandafter\ifx\csname natexlab\endcsname\relax\def\natexlab#1{#1}\fi
\expandafter\ifx\csname bibnamefont\endcsname\relax
  \def\bibnamefont#1{#1}\fi
\expandafter\ifx\csname bibfnamefont\endcsname\relax
  \def\bibfnamefont#1{#1}\fi
\expandafter\ifx\csname citenamefont\endcsname\relax
  \def\citenamefont#1{#1}\fi
\expandafter\ifx\csname url\endcsname\relax
  \def\url#1{\texttt{#1}}\fi
\expandafter\ifx\csname urlprefix\endcsname\relax\def\urlprefix{URL }\fi
\providecommand{\bibinfo}[2]{#2}
\providecommand{\eprint}[2][]{\url{#2}}

\bibitem[{\citenamefont{Carlo et~al.}(2007)\citenamefont{Carlo, Irmia,
  Tompkins, and Toner}}]{CarloToner}
\bibinfo{author}{\bibfnamefont{D.} \bibnamefont{Di Carlo}},
  \bibinfo{author}{\bibfnamefont{D.}~\bibnamefont{Irmia}},
  \bibinfo{author}{\bibfnamefont{R.~G.} \bibnamefont{Tompkins}},
  \bibnamefont{and} \bibinfo{author}{\bibfnamefont{M.}~\bibnamefont{Toner}},
  \bibinfo{journal}{PNAS} \textbf{\bibinfo{volume}{104}},
  \bibinfo{pages}{18892} (\bibinfo{year}{2007}).
\bibinfo{author}{\bibfnamefont{D.} \bibnamefont{Di Carlo}},
  \bibinfo{author}{\bibfnamefont{J.~F.} \bibnamefont{Edd}},
  \bibinfo{author}{\bibfnamefont{K.~J.} \bibnamefont{Humphry}},
  \bibinfo{author}{\bibfnamefont{H.~A.} \bibnamefont{Stone}}, \bibnamefont{and}
  \bibinfo{author}{\bibfnamefont{M.}~\bibnamefont{Toner}},
  \bibinfo{journal}{Phys. Rev. Lett.} \textbf{\bibinfo{volume}{102}},
  \bibinfo{pages}{094503} (\bibinfo{year}{2009}).

\bibitem[{\citenamefont{Seinfeld and Pandis}(1998)}]{Seinfeld98}
\bibinfo{author}{\bibfnamefont{J.~H.} \bibnamefont{Seinfeld}} \bibnamefont{and}
  \bibinfo{author}{\bibfnamefont{S.~N.} \bibnamefont{Pandis}},
  \emph{\bibinfo{title}{Atmospheric {C}hemistry and {P}hysics}}
  (\bibinfo{publisher}{Whiley, {N}ew {Y}ork}, \bibinfo{year}{1998}).


\bibitem[{\citenamefont{Maxey}(1987)}]{Max87}
E.g.,
\bibinfo{author}{\bibfnamefont{M.~R.} \bibnamefont{Maxey}},
  \bibinfo{journal}{Physics of Fluids} \textbf{\bibinfo{volume}{30}},
  \bibinfo{pages}{1915} (\bibinfo{year}{1987}).

\bibitem[{\citenamefont{Pasquero et~al.}(2003)\citenamefont{Pasquero,
  Provenzale, and Spiegel}}]{Gravity}
\bibinfo{author}{\bibfnamefont{C.}~\bibnamefont{Pasquero}},
  \bibinfo{author}{\bibfnamefont{A.}~\bibnamefont{Provenzale}},
  \bibnamefont{and} \bibinfo{author}{\bibfnamefont{E.~A.}
  \bibnamefont{Spiegel}}, \bibinfo{journal}{Phys. Rev. Lett.}
  \textbf{\bibinfo{volume}{91}}, \bibinfo{pages}{054502}
  (\bibinfo{year}{2003}).
\bibinfo{author}{\bibfnamefont{C.}~\bibnamefont{Escauriaza}} \bibnamefont{and}
  \bibinfo{author}{\bibfnamefont{F.}~\bibnamefont{Sotiropoulos}},
  \bibinfo{journal}{J. Fluid Mech.} \textbf{\bibinfo{volume}{641}},
  \bibinfo{pages}{169} (\bibinfo{year}{2009}).

\bibitem[{\citenamefont{Sapsis and Haller}(2009)}]{SaH}
\bibinfo{author}{\bibfnamefont{T.}~\bibnamefont{Sapsis}} \bibnamefont{and}
  \bibinfo{author}{\bibfnamefont{G.}~\bibnamefont{Haller}},
  \bibinfo{journal}{Journal of the Atmospheric Sciences}
  \textbf{\bibinfo{volume}{66}}, \bibinfo{pages}{2481} (\bibinfo{year}{2009}).
\bibinfo{author}{\bibfnamefont{T.}~\bibnamefont{Sapsis}} \bibnamefont{and}
  \bibinfo{author}{\bibfnamefont{G.}~\bibnamefont{Haller}},
  \bibinfo{journal}{CHAOS} \textbf{\bibinfo{volume}{20}},
  \bibinfo{pages}{017515} (\bibinfo{year}{2010}).

\bibitem[{\citenamefont{Vilela and Motter}(2007)}]{VilelaMotter07}
\bibinfo{author}{\bibfnamefont{R.~D.} \bibnamefont{Vilela}} \bibnamefont{and}
  \bibinfo{author}{\bibfnamefont{A.~E.} \bibnamefont{Motter}},
  \bibinfo{journal}{Phys. Rev. Lett.} \textbf{\bibinfo{volume}{99}},
  \bibinfo{pages}{264101} (\bibinfo{year}{2007}).

\bibitem[{\citenamefont{Benczik et~al.}(2002)\citenamefont{Benczik, Toroczkai,
  and T\'el}}]{BTT02}
\bibinfo{author}{\bibfnamefont{I.~J.} \bibnamefont{Benczik}},
  \bibinfo{author}{\bibfnamefont{Z.}~\bibnamefont{Toroczkai}},
  \bibnamefont{and} \bibinfo{author}{\bibfnamefont{T.}~\bibnamefont{T\'el}},
  \bibinfo{journal}{Phys. Rev. Lett.} \textbf{\bibinfo{volume}{89}},
  \bibinfo{pages}{164501} (\bibinfo{year}{2002}).


\bibitem[{\citenamefont{Schwabe et~al.}(1996)\citenamefont{Schwabe, Hintz, and
  Frank}}]{SHF96}
\bibinfo{author}{\bibfnamefont{D.}~\bibnamefont{Schwabe}},
  \bibinfo{author}{\bibfnamefont{P.}~\bibnamefont{Hintz}}, \bibnamefont{and}
  \bibinfo{author}{\bibfnamefont{S.}~\bibnamefont{Frank}},
  \bibinfo{journal}{Microgravity Sci. Technol.} \textbf{\bibinfo{volume}{9}},
  \bibinfo{pages}{163} (\bibinfo{year}{1996}).

\bibitem[{\citenamefont{Tanaka et~al.}(2006)\citenamefont{Tanaka, Kawamura,
  Ueno, and Schwabe}}]{TanakaSchwabe06}
\bibinfo{author}{\bibfnamefont{S.}~\bibnamefont{Tanaka}},
  \bibinfo{author}{\bibfnamefont{H.}~\bibnamefont{Kawamura}},
  \bibinfo{author}{\bibfnamefont{I.}~\bibnamefont{Ueno}}, \bibnamefont{and}
  \bibinfo{author}{\bibfnamefont{D.}~\bibnamefont{Schwabe}},
  \bibinfo{journal}{Physics of Fluids} \textbf{\bibinfo{volume}{18}}
  (\bibinfo{year}{2006}).

\bibitem[{\citenamefont{Schwabe et~al.}(2007)\citenamefont{Schwabe, Mizev,
  Udhayasankar, and Tanaka}}]{SMT07}
\bibinfo{author}{\bibfnamefont{D.}~\bibnamefont{Schwabe}},
  \bibinfo{author}{\bibfnamefont{A.~I.} \bibnamefont{Mizev}},
  \bibinfo{author}{\bibfnamefont{M.}~\bibnamefont{Udhayasankar}},
  \bibnamefont{and} \bibinfo{author}{\bibfnamefont{S.}~\bibnamefont{Tanaka}},
  \bibinfo{journal}{Physics of Fluids} \textbf{\bibinfo{volume}{19}},
  \bibinfo{pages}{072102} (\bibinfo{year}{2007}).

\bibitem[{\citenamefont{Pushkin
  et~al.}(2010{\natexlab{a}})\citenamefont{Pushkin, Melnikov, and
  Shevtsova}}]{PMS10}
\bibinfo{author}{\bibfnamefont{D.}~\bibnamefont{Pushkin}},
  \bibinfo{author}{\bibfnamefont{D.}~\bibnamefont{Melnikov}}, \bibnamefont{and}
  \bibinfo{author}{\bibfnamefont{V.}~\bibnamefont{Shevtsova}}, in
  \emph{\bibinfo{booktitle}{5th {M}eeting of the {I}nternational {M}arangoni
  {A}ssociation}} (\bibinfo{address}{Florence},
  \bibinfo{year}{2010}{\natexlab{a}}).


\bibitem[{\citenamefont{Hide and Mason}(1975)}]{HM75}
\bibinfo{author}{\bibfnamefont{R.}~\bibnamefont{Hide}} \bibnamefont{and}
  \bibinfo{author}{\bibfnamefont{P.~J.} \bibnamefont{Mason}},
  \bibinfo{journal}{Adv. Phys.} \textbf{\bibinfo{volume}{24}},
  \bibinfo{pages}{47} (\bibinfo{year}{1975}).

\bibitem[{\citenamefont{Lopez et~al.}(2004)\citenamefont{Lopez, Marques, Hirsa,
  and Miraghaie}}]{Lopez04}
\bibinfo{author}{\bibfnamefont{J.~M.} \bibnamefont{Lopez}},
  \bibinfo{author}{\bibfnamefont{F.}~\bibnamefont{Marques}},
  \bibinfo{author}{\bibfnamefont{A.~H.} \bibnamefont{Hirsa}}, \bibnamefont{and}
  \bibinfo{author}{\bibfnamefont{R.}~\bibnamefont{Miraghaie}},
  \bibinfo{journal}{J. Fluid Mech.} \textbf{\bibinfo{volume}{502}},
  \bibinfo{pages}{99} (\bibinfo{year}{2004}).

\bibitem[{\citenamefont{Alvarez et~al.}(2002)\citenamefont{Alvarez, Zalc,
  Shinbrot, Arratia, and Muzzio}}]{AlvarezShinbrotMuzzio02}
\bibinfo{author}{\bibfnamefont{M.~M.} \bibnamefont{Alvarez}},
  \bibinfo{author}{\bibfnamefont{J.~M.} \bibnamefont{Zalc}},
  \bibinfo{author}{\bibfnamefont{T.}~\bibnamefont{Shinbrot}},
  \bibinfo{author}{\bibfnamefont{P.~E.} \bibnamefont{Arratia}},
  \bibnamefont{and} \bibinfo{author}{\bibfnamefont{F.~J.}
  \bibnamefont{Muzzio}}, \bibinfo{journal}{AIChE Journal}
  \textbf{\bibinfo{volume}{48}}, \bibinfo{pages}{2135} (\bibinfo{year}{2002}).

\bibitem[{\citenamefont{Preisser et~al.}(1983)\citenamefont{Preisser, Schwabe,
  and Scharmann}}]{PSS83}
\bibinfo{author}{\bibfnamefont{F.}~\bibnamefont{Preisser}},
  \bibinfo{author}{\bibfnamefont{D.}~\bibnamefont{Schwabe}}, \bibnamefont{and}
  \bibinfo{author}{\bibfnamefont{A.}~\bibnamefont{Scharmann}},
  \bibinfo{journal}{J. Fluid Mech.} \textbf{\bibinfo{volume}{126}},
  \bibinfo{pages}{545} (\bibinfo{year}{1983}).

\bibitem[{\citenamefont{Scriven and Sternling}(1960)}]{ScrivenSternling60}
\bibinfo{author}{\bibfnamefont{L.~E.} \bibnamefont{Scriven}} \bibnamefont{and}
  \bibinfo{author}{\bibfnamefont{C.~V.} \bibnamefont{Sternling}},
  \bibinfo{journal}{Nature} \textbf{\bibinfo{volume}{187}},
  \bibinfo{pages}{186} (\bibinfo{year}{1960}).

\bibitem[{\citenamefont{Scriven and Sternling}(1960)}]{SupplMat}
Two supplementary movies. Other experimental movies are available free of charge following the links in \cite{SMT07}. 



\bibitem[{\citenamefont{Burgisser and Bergantz}(2002)}]{BB02}
\bibinfo{author}{\bibfnamefont{A.}~\bibnamefont{Burgisser}} \bibnamefont{and}
  \bibinfo{author}{\bibfnamefont{G.~W.} \bibnamefont{Bergantz}},
  \bibinfo{journal}{Earth and Planetary Science Letters}
  \textbf{\bibinfo{volume}{202}}, \bibinfo{pages}{405} (\bibinfo{year}{2002}).
\bibinfo{author}{\bibfnamefont{K.}~\bibnamefont{Sefiane}},
  \bibinfo{author}{\bibfnamefont{J.~R.} \bibnamefont{Moffat}},
  \bibinfo{author}{\bibfnamefont{O.~K.} \bibnamefont{Matar}}, \bibnamefont{and}
  \bibinfo{author}{\bibfnamefont{R.~V.} \bibnamefont{Craster}},
  \bibinfo{journal}{Applied Physics Letters} \textbf{\bibinfo{volume}{93}},
  \bibinfo{pages}{074103} (\bibinfo{year}{2008}).




\bibitem[{\citenamefont{Maxey and Riley}(1983)}]{MREq}
\bibinfo{author}{\bibfnamefont{M.~R.} \bibnamefont{Maxey}} \bibnamefont{and}
  \bibinfo{author}{\bibfnamefont{J.~J.} \bibnamefont{Riley}},
  \bibinfo{journal}{Physics of Fluids} \textbf{\bibinfo{volume}{26}},
  \bibinfo{pages}{883} (\bibinfo{year}{1983}).
\bibinfo{author}{\bibfnamefont{E.~E.} \bibnamefont{Michaelides}},
  \bibinfo{journal}{Journal of Fluid Engineering}
  \textbf{\bibinfo{volume}{119}}, \bibinfo{pages}{233} (\bibinfo{year}{1997}).
\bibinfo{author}{\bibfnamefont{M.~R.} \bibnamefont{Maxey}},
  \bibinfo{journal}{Phys. Fluids}
  \textbf{\bibinfo{volume}{30}}\bibinfo{number}{(7)}, \bibinfo{pages}{1915} (\bibinfo{year}{1987}).



\bibitem[{\citenamefont{Shevtsova et~al.}(2009)\citenamefont{Shevtsova,
  Melnikov, and Nepomnyashchy}}]{ShevMelNep09}
\bibinfo{author}{\bibfnamefont{V.}~\bibnamefont{Shevtsova}},
  \bibinfo{author}{\bibfnamefont{D.~E.} \bibnamefont{Melnikov}},
  \bibnamefont{and}
  \bibinfo{author}{\bibfnamefont{A.}~\bibnamefont{Nepomnyashchy}},
  \bibinfo{journal}{Phys. Rev. Lett.} \textbf{\bibinfo{volume}{102}},
  \bibinfo{pages}{134503} (\bibinfo{year}{2009}).

\bibitem[{\citenamefont{Melnikov et~al.}(2011)\citenamefont{Melnikov, Pushkin,
  and Shevtsova}}]{MelPushShev10}
\bibinfo{author}{\bibfnamefont{D.}~\bibnamefont{Melnikov}},
  \bibinfo{author}{\bibfnamefont{D.}~\bibnamefont{Pushkin}}, \bibnamefont{and}
  \bibinfo{author}{\bibfnamefont{V.}~\bibnamefont{Shevtsova}},
  \bibinfo{journal}{European Physical Journal - SpecialTopics}  \textbf{\bibinfo{volume}{192}},
  \bibinfo{pages}{29} (\bibinfo{year}{2011}).

\bibitem[{\citenamefont{Pushkin
  et~al.}(2011{\natexlab{b}})\citenamefont{Pushkin, Melnikov, and
  Shevtsova}}]{PMS_LongPaper10}
\bibinfo{author}{\bibfnamefont{D.}~\bibnamefont{Pushkin}},
  \bibinfo{author}{\bibfnamefont{D.}~\bibnamefont{Melnikov}}, \bibnamefont{and}
  \bibinfo{author}{\bibfnamefont{V.}~\bibnamefont{Shevtsova}}
  (\bibinfo{year}{2011}{\natexlab{b}}), \bibinfo{note}{manuscript in
  preparation}.


\bibitem[{\citenamefont{Schwabe and Mizev}(2011)}]{SM10}
\bibinfo{author}{\bibfnamefont{D.}~\bibnamefont{Schwabe}} \bibnamefont{and}
  \bibinfo{author}{\bibfnamefont{A.}~\bibnamefont{Mizev}},
\bibinfo{journal}{European Physical Journal - Special Topics} \textbf{\bibinfo{volume}{192}},
  \bibinfo{pages}{13} (\bibinfo{year}{2011}).


\bibitem[{\citenamefont{Ott}(1993)}]{Ott93}
\bibinfo{author}{\bibfnamefont{E.}~\bibnamefont{Ott}},
  \emph{\bibinfo{title}{Chaos in dynamical systems}}
  (\bibinfo{publisher}{Cambridge University Press}, \bibinfo{year}{1993}).

\end{thebibliography}

\end{document}